\documentclass{nature}

\usepackage{graphicx}

\spacing{1.2}

\begin{document}

\newcommand{\Ha}{\ensuremath{{\rm H\alpha}}}

{\newpage\spacing{1}\setlength{\parskip}{12pt}%
  {\Large\bfseries\noindent\sloppy \textsf{
      High star formation rates as the origin of turbulence in early and modern disk galaxies } \par}}


{\spacing{1}\noindent\sloppy Green, Andrew W.$^{1}$; Glazebrook, Karl$^{1}$;
  McGregor, Peter J.$^{2}$; Abraham, Roberto G.$^{3}$; Poole, Gregory
  B.$^{1}$; Damjanov, Ivana$^{3}$; McCarthy, Patrick J.$^{4}$,
  Colless, Matthew$^{5}$ \& Sharp, Robert G.$^{5}$}

{\spacing{1}
\begin{affiliations}
 \item Centre for Astrophysics and Supercomputing, Swinburne
   University, PO Box 218, Hawthorn, VIC 3122, Australia
 \item Research School of Astronomy and Astrophysics, Australian
   National University, Cotter Rd, Weston, ACT 2611, Australia
 \item Department of Astronomy and Astrophysics, University of
   Toronto, 50 St George St, Toronto, ON M5S3H4, Canada
 \item Observatories of the Carnegie Institution of Washington, 813
   Santa Barbara Street, Pasadena, CA 91101, USA
 \item Australian Astronomical Observatory, PO Box 296, Epping, NSW
   1710, Australia
\end{affiliations}
}


\begin{abstract}

  High
  spatial and spectral resolution observations of star formation and
  kinematics in early galaxies have shown that two-thirds are massive
  rotating disk galaxies\cite{2006Natur.442..786G,
    2009ApJ...706.1364F, 2008Natur.455..775S, 2009A&A...504..789E,
    2010MNRAS.404.1247J} with the remainder being less massive
  non-rotating objects\cite{2009A&A...504..789E, 2006A&A...455..107F,
    2009ApJ...706.1364F, 2009ApJ...697.2057L, 2010MNRAS.401.1657L}.
  The line of sight averaged velocity dispersions are typically five
  times higher than in today's disk galaxies. This has suggested that
  gravitationally-unstable, gas-rich disks in the early Universe are
  fuelled by cold, dense accreting gas flowing along cosmic filaments
  and penetrating hot galactic gas halos\cite{2009Natur.457..451D,
    2008ApJ...688...67E}.  However these accreting flows have not been
  observed\cite{2010arXiv1003.0679S}, and cosmic accretion cannot
  power the observed level of
  turbulence\cite{2010ApJ...712..294E}. Here we report on a new sample
  of rare high-velocity-dispersion disk galaxies we have discovered in
  the nearby Universe where cold accretion is unlikely to drive their
  high star-formation rates.  We find that the velocity dispersion is
  most fundamentally correlated with their star-formation rates, and
  not their mass nor gas fraction, which leads to a new picture where
  star formation itself is the energetic driver of galaxy disk
  turbulence at all cosmic epochs.

\end{abstract}




Understanding how these different kinematic states of star-forming
galaxies fit together is complicated by selection, surface-brightness
and angular-resolution effects. Particularly at high redshift,
resolution is a major limitation. The resolution gain of adaptive
optics has enabled kinematic observations of early disks but not all
observations, even within a particular survey, have taken advantage of
adaptive optics\cite{2009ApJ...706.1364F, 2009A&A...504..789E,
  2010MNRAS.401.1657L}. Integral-Field Spectroscopy (IFS) (mainly of
the H$\alpha$ emission line, which traces star formation and
kinematics) has shallower surface-brightness limits than integrated
spectroscopy. Kinematic observations are generally limited to the
brightest (the characteristic Schechter luminosity L* or
brighter\cite{2010MNRAS.404.1247J}) high-redshift galaxies. Surface
brightness is also a strong function of redshift (proportional to
$(1+z)^4$), and optical pass-bands commonly used for imaging sample
the restframe-UV. These effects often complicate comparisons between
the early and modern Universe. Finally, local IFS comparison samples
are often selected from small volumes and non-uniformly.


To quantify how these difficulties might affect previous results, we
turned to the well-studied Sloan Digital Sky Survey (SDSS) to put kinematic
galaxy states in the context of a large, uniformly-selected sample. We
undertook the first IFS observations of 65
star-forming\cite{2004MNRAS.351.1151B} galaxies at redshift $z \sim
0.1$. As active galactic nuclei interfere with \Ha\ emission as a star
formation tracer, they have been excluded. (Further details of the
selection criteria are in the Supplementary Information). Observed
galaxies have \Ha\ luminosities of $10^{40.7}$ to $10^{42.6}$ erg/s
and median of $10^{41.9}$ erg/s and stellar mass range and median of
$10^{9.1}$ to $10^{10.9}$ and $10^{10.3}$ solar masses,
respectively\cite{2003MNRAS.341...33K}. Because only bright (which we
shall define here as $L_\Ha > 10^{42}$erg/s) objects are typically
detected at $z \sim 2$, such objects make up half our observations
despite representing only 3.2\% of SDSS galaxies which otherwise meet
our criteria. A broad selection illustrates the impact of surface
brightness and luminosity on our results and those reported for the
high-redshift Universe.


Chosen galaxies were observed using the integral-field spectrographs
SPIRAL\cite{2006SPIE.6269E..14S} on the 3.9m Anglo-Australian
Telescope or WiFeS\cite{2007Ap&SS.310..255D} on the ANU 2.3m
telescope. Median seeing of 1.3'' corresponds to a median spatial
resolution of 2.3 kpc and the field of view is 10--40 kpc. This is
closely matched to high-redshift samples observed with adaptive
optics, but with better spectral resolution ($ \lambda / \Delta
\lambda = 7000$ to 11,500).  Following standard methods, we fit a Gaussian
profile to the \Ha\ emission-line spectrum at each spatial position in
the reduced data cube. The free parameters of the fit are velocity
dispersion (width, corrected for instrumental broadening), flux
(height), and velocity (position). The velocity dispersion ($\sigma$)
is a simple, model-independent measure of the line-of-sight
kinematics of a galaxy. Although there are many ways to measure it, we adopt the
flux-weighted local mean definition commonly used at high redshift
\cite{2009A&A...504..789E,2009ApJ...697.2057L,2010MNRAS.401.1657L}:
\begin{equation}
  \label{eq:sigma_mean}
  \sigma_{\rm m} = \frac{\sum{\sigma_{\rm pix} f_{\rm pix}}}{\sum{f_{\rm pix}}}
\end{equation}
where $\sigma_{\rm pix}$ is the standard deviation of the Gaussian fit
to the \Ha\ emission line in each spatial pixel, and $f_{\rm pix}$ is
the \Ha\ flux in that spatial pixel. This quantity measures the
intrinsic velocity dispersion of the HII star forming regions and
their random motions independent of large-scale systematic motions,
such as rotation or orbital motion in a merger. Apparent velocity
dispersion can also peak where systematic motions change abruptly,
such as the bright central regions of a disk galaxy where the velocity
curve is steepest. We demonstrate that this does not affect our
results in the Supplementary Information.


Unexpectedly, we find many disk-like velocity fields (Figure 1) with
$\sigma_m > 50$ km/s (high dispersion) among the galaxies with $L_\Ha
> 10^{42}$ erg/s (high luminosity), very similar values to those at
high redshift.  We use two simple, qualitative criteria common to
high-redshift analyses\cite{2006A&A...455..107F} to separate disks
from non-disks based on their velocity and velocity-dispersion fields:
(1) the velocity field must have a typical ``spider diagram'' shape,
and (2) the velocity dispersion must be centrally peaked and fairly
symmetric. Using these criteria all but six of our 17 objects at high
dispersion and high luminosity are kinematic disks. The broad-band
SDSS images also show most are disk galaxies, often with distinct
bulges and disk components. This agrees with the picture that most of
the star formation in the local Universe occurs in
disks\cite{2008A&A...482..507J} and those disks forming stars most
rapidly are kinematically hot, just as in the early Universe.


We compare the distribution of our galaxies in $\sigma_m$ and \Ha\
luminosity (shown in Figure 2) with several existing data sets from
the early\cite{2009ApJ...697.2057L, 2010MNRAS.401.1657L,
  2009A&A...504..789E} and modern\cite{2008MNRAS.390..466E,
  2010MNRAS.401.2113E, 1974ApJS...27..415T} Universe where $\sigma_m$
has been measured in a comparable way (a summary of these data sets
appears in the Supplementary Information). To check that
surface-brightness dimming with redshift does not affect this
comparison, we have applied redshift dimming (to $z=2.2$) and a
surface-brightness limit ($>10^{16}$erg/s/cm$^2$/arcsec$^2$, half that
reported for OSIRIS\cite{2009ApJ...697.2057L}) to our data and
recomputed $\sigma_m$. We find this affects our measure of $\sigma_m$
by only 5--10 km/s. Furthermore, all of these comparison data sets
fall within the stellar mass range of our data. Remarkably, $\sigma_m$
appears to correlate with \Ha\ luminosity above $L_\Ha = 10^{42}$
erg/s independent of redshift.


The H$\alpha$ luminosity is directly correlated with the star-formation
rate\cite{1998ARA&A..36..189K} (SFR). However, dust extinction in the
galaxy can obscure some of this light, reducing the inferred
SFR. Because dust-correction methods differ between authors, we use
the observed quantity $L_\Ha$ to provide a simple, direct comparison
between samples. In our sample, the mean flux ratio, ${\rm
  H\alpha}/{\rm H\beta} = 4.13$ (based on SDSS fibre spectra)
corresponds to a typical extinction at H$\alpha$ of $A_{\Ha}=0.78$
mag\cite{1992ApJ...395..130P}. This correction has been included in
the axis across the top of Figure 2 to give a rough scale to the
star-formation rate for these galaxies.


Although primordial galaxies have been interpreted as turbulent
gas-rich disks\cite{2009ApJ...706.1364F, 2008ApJ...688...67E}, our
results suggest gas density does not drive velocity
dispersion. Applying the Kennicutt-Schmidt
Law\cite{1998ApJ...498..541K} to invert our star-formation surface
densities into gas surface densities, we can estimate the gas-mass
fraction (relative to stars and gas) for these galaxies to range from
0.05 to 0.7 (median 0.18). The distributions of gas-fractions between
high- and low-dispersion galaxies do not differ significantly, nor do
the gas depletion timescales of 3--6 billion years. The wide range in
gas fraction over a small range in luminosity is due to the wide range
in stellar mass of these galaxies---stellar mass does not correlate
with \Ha\ luminosity. Figure 2 shows that there is no correlation
between stellar mass and velocity dispersion. This implies
star-formation rate is the important variable driving different
velocity dispersions, not mass nor gas fraction, at all redshifts.


These high-dispersion galactic disks in the local Universe are an
unexpected find. Continued detailed followup of these objects will
shed more light on similar objects seen at high redshift. Cold-flow
accretion is unlikely to be the origin of the high velocity
dispersions as this mechanism is expected\cite{2006MNRAS.368....2D} to
shut down rapidly for galaxies at $z<2$. The high velocity dispersions
in low-redshift galaxies then raise a question of whether cold-flow
accretion is the appropriate mechanism for high velocity dispersions
at high-redshifts.  Recent absorption-line measurements have found no
evidence for ubiquitous cold flows in the high-redshift
Universe\cite{2010arXiv1003.0679S}. Our results suggest star formation
itself powers the turbulence through energetic feedback. Indeed,
simulations show that supernovae resulting from star formation can
drive high velocity dispersions in the interstellar
medium\cite{2006ApJ...638..797D}.


It remains necessary to provide a mechanism to fuel the high
star-formation rates of these galaxies.  While the kinematics of our
sample are not merger-dominated, there is evidence in the optical
morphologies for minor-merger features such as small, close companions
or tidal tails, many of which would be missed by current observations
at high-redshift (e.g. panels f and j in Figure 1). Fresh gas brought
in by these minor mergers may drive the high star-formation
rates. These mergers could also inflate the velocity dispersions of
the disks. The situation could be similar in the early Universe; the
minor merger mechanism has been commonly invoked to drive the dramatic
size evolution of red galaxies\cite{2009Natur.460..694G}, requiring
minor mergers to occur frequently. Some of our objects also show
companions at large distances with luminosity ratios between 1:2 to
1:3. This suggests gravitational tides could be compressing gas and
inducing star-formation. Such a `fly-by' mechanism would be
considerably more important in the high-redshift
Universe\cite{2005ApJ...632...49M} (because of the high merger rate)
and such objects could be massive, yet faint at optical wavelengths.
Further study of the high-dispersion local galaxies is warranted to
determine the physical mechanisms operating. These objects serve as
rare reminders of the massive star formation that was so prevalent in
the Universe 10 billion years ago.




\newcommand\lt{$<$}
\newcommand\gt{$>$}

\newcommand\aj{Astron. J.}%
\newcommand\actaa{Acta Astron.}%
\newcommand\araa{ARA\&A}%
\newcommand\apj{Astrophys. J.}%
\newcommand\apjl{Astrophys. J. L.J}%
\newcommand\apjs{Astrophys. J. S.}%
\newcommand\ao{Appl.~Opt.}%
\newcommand\apss{Astrophys.\& Space Sci.}%
\newcommand\aap{Astron. Astrophys.}%
\newcommand\aapr{A\&A~Rev.}%
\newcommand\aaps{A\&AS}%
\newcommand\azh{AZh}%
\newcommand\baas{BAAS}%
\newcommand\caa{Chinese Astron. Astrophys.}%
\newcommand\cjaa{Chinese J. Astron. Astrophys.}%
\newcommand\icarus{Icarus}%
\newcommand\jcap{J. Cosmology Astropart. Phys.}%
\newcommand\jrasc{JRASC}%
\newcommand\memras{MmRAS}%
\newcommand\mnras{Mon. Not. R. Astron. Soc.}%
\newcommand\na{New A}%
\newcommand\nar{New A Rev.}%
\newcommand\pra{Phys.~Rev.~A}%
\newcommand\prb{Phys.~Rev.~B}%
\newcommand\prc{Phys.~Rev.~C}%
\newcommand\prd{Phys.~Rev.~D}%
\newcommand\pre{Phys.~Rev.~E}%
\newcommand\prl{Phys.~Rev.~Lett.}%
\newcommand\pasa{PASA}%
\newcommand\pasp{PASP}%
\newcommand\pasj{PASJ}%
\newcommand\qjras{QJRAS}%
\newcommand\rmxaa{Rev. Mexicana Astron. Astrofis.}%
\newcommand\skytel{S\&T}%
\newcommand\solphys{Sol.~Phys.}%
\newcommand\sovast{Soviet~Ast.}%
\newcommand\ssr{Space~Sci.~Rev.}%
\newcommand\zap{ZAp}%
\newcommand\nat{Nature}%
\newcommand\iaucirc{IAU~Circ.}%
\newcommand\aplett{Astrophys.~Lett.}%
\newcommand\apspr{Astrophys.~Space~Phys.~Res.}%
\newcommand\bain{Bull.~Astron.~Inst.~Netherlands}%
\newcommand\fcp{Fund.~Cosmic~Phys.}%
\newcommand\gca{Geochim.~Cosmochim.~Acta}%
\newcommand\grl{Geophys.~Res.~Lett.}%
\newcommand\jcp{J.~Chem.~Phys.}%
\newcommand\jgr{J.~Geophys.~Res.}%
\newcommand\jqsrt{J.~Quant.~Spec.~Radiat.~Transf.}%
\newcommand\memsai{Mem.~Soc.~Astron.~Italiana}%
\newcommand\nphysa{Nucl.~Phys.~A}%
\newcommand\physrep{Phys.~Rep.}%
\newcommand\physscr{Phys.~Scr}%
\newcommand\planss{Planet.~Space~Sci.}%
\newcommand\procspie{Proc.~SPIE}%

\bibliographystyle{naturemag}
\bibliography{bibliographies}



\begin{addendum}
  
\item[Supplementary Information] is linked to the online version of
  the paper at www.nature.com/nature

\item[Acknowledgements]A.W.G and K.G. acknowledges financial support
  from the Australian Research Council.
  A.W.G. acknowledges a special scholarship from the Chancellery of
  the Swinburne University of Technology. We wish to thank the staff
  of the Anglo-Australian Observatory and the staff of the ANU 2.3m
  telescope for their support of these observations.
 \item[Author Contributions] K.G. oversaw the project. A.W.G.,
   K.G., I.D., P.J. McGregor, G.P. and R.G.S. collected the data at the
   telescope. A.W.G. completed the data reduction with help from P.J. McGregor
   and R.G.S. M.C. kindly provided observing time. A.W.G. and K.G. analysed the
   data and wrote the paper. All authors provided extensive suggestions
   and comments at each stage of the project.
 \item[Competing Interests] The authors declare that they have no
competing financial interests.
 \item[Correspondence] Correspondence and requests for materials
should be addressed to A.W.G. (via email: agreen@astro.swin.edu.au).
\end{addendum}


\pagestyle{empty}

  \noindent\begin{minipage}{95mm}
  \includegraphics[width=89mm]{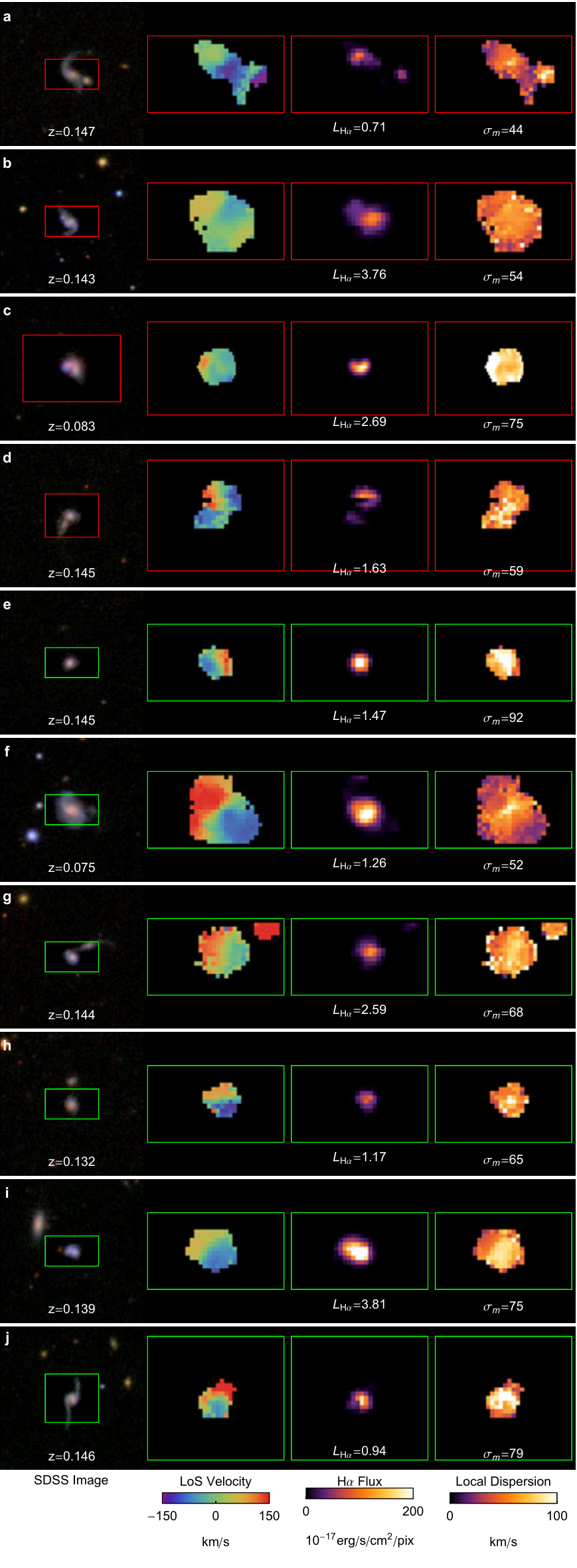}
  \end{minipage}
  \begin{minipage}[ct]{80mm}
\sffamily \footnotesize
    {\bfseries Figure 1: Kinematic Pictures of Local galaxies}\\
    All panels show the SDSS broadband image (red is $\sim$750nm,
    green~$\sim$625nm and blue~$\sim$475nm), line-of-sight velocity
    map, \Ha\ narrowband image, and line-of-sight velocity dispersion
    map. All are scaled identically according to the scale at the
    bottom of the figure. The maps are masked when the emission-line
    fitting breaks down, typically at a $4\sigma$ significance
    detection threshold. The green or red squares outline the field of
    view of the instrument in the images; red corresponds to non-disk
    objects and green to disk-like object. The broadband imaging of
    ({\bf a, b,} and {\bf d}) show clear merger activity, while ({\bf
      c}) may have a merger axis close to the line of sight. Panel
    {\bf (a)} shows a relatively small velocity dispersion, despite
    being a clear merger, suggesting that dispersion alone cannot
    distinguish mergers.  Panels {\bf e-j} are all objects with high
    dispersions ($\sigma > 50$ km/s), but which are considered disks
    by our criteria. We emphasize these criteria are based on those
    typically employed on high-redshift
    observations\cite{2006A&A...455..107F}. ({\bf f}) is a clear disk,
    with both a blue disk and a red bulge component visible in the
    broadband image. ({\bf g}) shows a velocity gradient and a weak
    central dispersion peak. This galaxy's neighbour is 0.62 mag
    fainter, and has a similar radial velocity. The pair may
    ultimately become a major merger. {\bf (h)} also has a companion
    which may or may not be at the same redshift. {\bf (i)} shows two
    distinct clumps in the broadband image, indicative of a major
    merger, but still shows the clear kinematic characteristics of a
    disk. Finally {\bf (j)} shows what one might argue is the
    signature of a disk in the kinematic data, but is a clear merger,
    with long tidal tails in the image. With the strong
    surface-brightness dimming with redshift, these tails would likely
    be invisible in a similar galaxy in the early Universe, and a disk
    interpretation would be made.

  \end{minipage}

 \noindent
 \includegraphics[width=89mm]{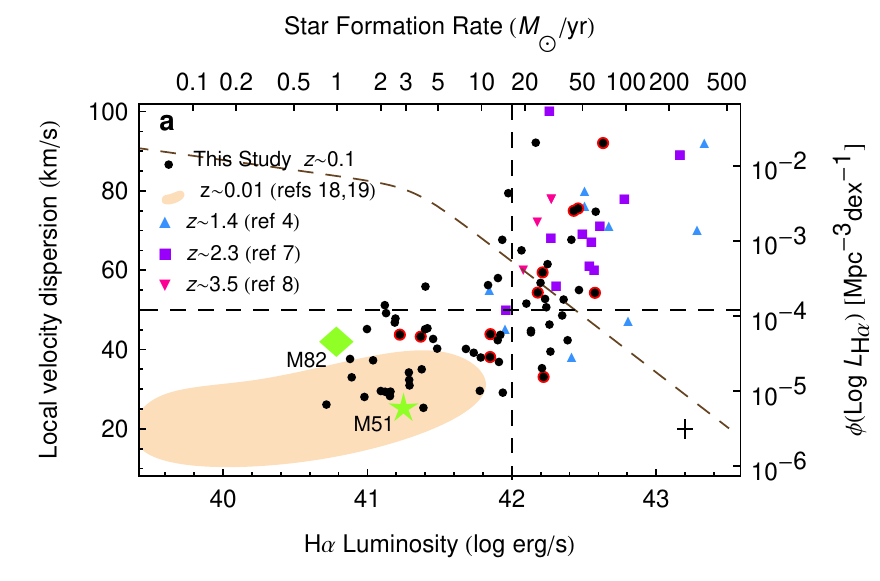}
 \includegraphics[width=89mm]{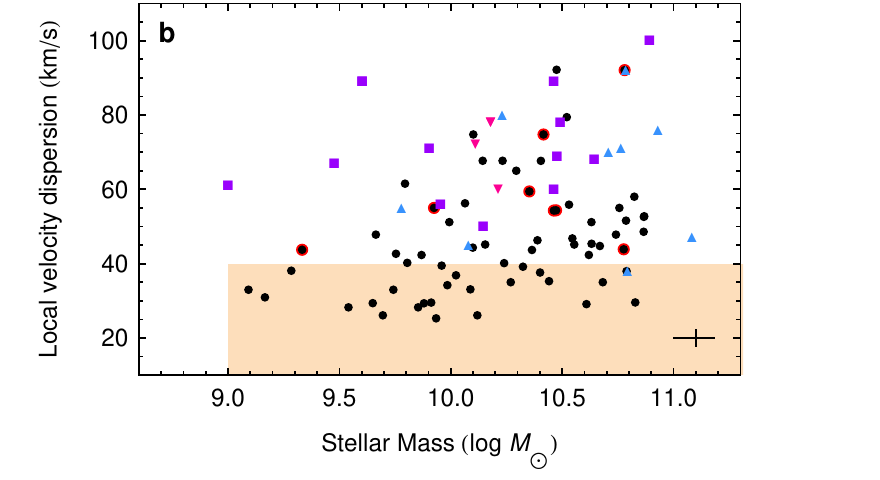}

\noindent
{\sffamily \footnotesize {\bfseries Figure 2: Distribution of galaxy velocity dispersion
     against \Ha\
     luminosity and stellar mass}\\
    Panel (a) shows the distribution of flux weighted mean velocity
    dispersion and \Ha\ luminosity or star-formation rate of galaxies
    in our sample and comparison samples at $z \sim 2$\cite{
      2009A&A...504..789E, 2009ApJ...697.2057L, 2010MNRAS.401.1657L}
    and $z < 0.01$\cite{ 2008MNRAS.390..466E,
      2010MNRAS.401.2113E}. The error bar in the lower right shows the
    combined, median errors. Velocity dispersion errors are typically
    $\pm$5--10\% (systematic) and $\pm$1--2 km/s (statistical), and
    luminosity errors $\pm$10\% (statistical). The variance and error
    in velocity dispersion are discussed in the Supplementary
    Information. Panel (b) shows the same, but plotted against stellar
    mass instead of \Ha\ luminosity. The points are labeled according
    to the key in panel (a).  The comparison samples are described in
    the Supplementary Information. The positions of local galaxies
    M51\cite{1974ApJS...27..415T} and M82\cite{1999ApJ...523..575L}
    are labeled. Dashed lines at $\sigma_m = 50$ km/s and $L_\Ha =
    10^{42}$ erg/s separate the different regimes described in the
    text. Velocity dispersion appears to correlate with \Ha\
    luminosity, but not mass. Non-disk galaxies are highlighted with
    red circles, and are not distinguishable by velocity dispersion or
    luminosity.  The \Ha\ luminosity is derived from the full aperture
    of our data cubes so does not suffer from the large aperture
    effects common to SDSS fibre spectra.\cite{2004MNRAS.351.1151B}
    The brown dashed line in ({\bf a}) is the \Ha\ luminosity function
    of SDSS galaxies\cite{2010MNRAS.405.2594G} with the scale on the
    right for comparison and shows the space density of star-forming
    galaxies declines sharply with higher luminosity.



\end{document}